\documentclass[prb,reprint,floatfix,amsmath,amssymb,showpacs]{revtex4-1}
\usepackage{amsfonts}%
\usepackage{mathrsfs,bbm}%
\usepackage{xcolor}
\usepackage{graphicx}%
  \DeclareGraphicsExtensions{.pdf,.eps,}
\usepackage{prettyref}%
	\newcommand{\pref}[1]{\prettyref{#1}}%
	\newrefformat{fig}{Fig.~\ref{#1}}%
	\newrefformat{tab}{Table~\ref{#1}}%
	\newrefformat{sec}{Sec.~\ref{#1}}%
	\newrefformat{eq}{Eq.~(\ref{#1})}%

\newcommand{\pwcond}{\texttt{PWcond}}
\newcommand{\pwscf}{\texttt{PWscf}}
\newcommand{\qe}{\textsc{Quantum~ESPRESSO}}
\newcommand{\de}{\partial}
\newcommand{\bra}[1]{\langle#1\,\lvert}
\newcommand{\ket}[1]{\rvert\,#1\rangle}
\newcommand{\bk}[2]{\langle#1\lvert\,#2\rangle}
\newcommand{\sw}[2]{\langle#2\lvert\,#1\,\rvert\,#2\rangle}
\newcommand{\mtx}[3]{\langle#1\lvert\,#2\,\rvert\,#3\rangle}
\newcommand{\bfr}{\ensuremath{\textbf{r}}}%
\newcommand{\bfR}{\ensuremath{\textbf{R}}}%
\newcommand{\bfk}{\ensuremath{\textbf{k}}}%
\newcommand{\ang}{\ensuremath{\textnormal{\AA}}}
\newcommand{\ry}{\ensuremath{{\mathrm{Ry}}}}%
\newcommand{\ev}{\ensuremath{{\mathrm{eV}}}}%
\newcommand{\hgo}{\ensuremath{{e^2/h}}}%
\newcommand{\ef}{\ensuremath{E_{\mathrm{F}}}}%
\newcommand{\echem}{\ensuremath{E_{\mathrm{chem}}}}%
\newcommand{\dco}{\ensuremath{d_{\textrm{C-O}}}}%
\newcommand{\dauc}{\ensuremath{d_{\textrm{Au-C}}}}%
\newcommand{\dwire}{\ensuremath{d_{\textrm{chain}}}}%
\let\dchain\dwire%
\newcommand{\dxz}{\ensuremath{d_{xz}}}%
\newcommand{\dyz}{\ensuremath{d_{yz}}}%
\newcommand{\sga}{\ensuremath{5\sigma_{\textrm{a}}}}%
\newcommand{\tps}{\ensuremath{2\pi^{\star}}}%
\newcommand{\nmmp}{\ensuremath{n_{mm'}^{I,\sigma}}}
\newcommand{\nmpm}{\ensuremath{n_{m'm}^{I,\sigma}}}
\newcommand{\pmmp}{\ensuremath{P_{mm'}^I}}

\newcommand{\fm}{\ensuremath{\varphi_{m}^{I}}}
\newcommand{\fmp}{\ensuremath{\varphi_{m'}^{I}}}
\newcommand{\vmmp}{\ensuremath{V_{mm'}^{I,\sigma}}}

\newcommand{\smm}{\ensuremath{\sum_{mm'}}}
\newcommand{\sis}{\ensuremath{\sum_{I,\sigma}}}
\newcommand{\vu}{\ensuremath{V_U^{\sigma}}}
\begin{document}


\title{Efficient DFT$+U$ calculations of ballistic electron transport:\
Application to Au monatomic chains with a CO impurity} 
\date{\today}
\author{Gabriele Sclauzero}
\email[Corresponding author: ]{gabriele.sclauzero@epfl.ch}
\altaffiliation[Present address: ]{%
Ecole Polytechnique F\'ed\'erale de Lausanne (EPFL), ITP-CSEA, CH-1015 Lausanne, Switzerland.}
\author{Andrea \surname{Dal Corso}}
\affiliation{%
International School for Advanced Studies (SISSA-ISAS), Via Bonomea 265, IT-34136 Trieste, Italy}
\affiliation{%
IOM-CNR Democritos, Via Bonomea 265, IT-34136 Trieste, Italy}

\begin{abstract}
  An efficient method for computing the Landauer-B\"uttiker conductance of an open quantum system within DFT$+U$ is presented.
  The Hubbard potential is included in electronic structure and transport calculations as a simple renormalization of the non-local pseudopotential coefficients by restricting the integration for the on-site occupations within the cutoff spheres of the pseudopotential.
  We apply the methodology to the case of an Au monatomic chain in presence of a CO molecule adsorbed on it.
  We show that the Hubbard $U$ correction removes the spurious magnetization in the pristine Au chain at the equilibrium spacing, as well as the unphysical contribution of $d$ electrons to the conductance, resulting in a single (spin-degenerate) transmission channel and a more realistic conductance of $1\,G_0$.
  We find that the conductance reduction due to CO adsorption is much larger for the atop site than for the bridge site, so that the general picture of electron transport in stretched Au chains given by the local density approximation remains valid at the equilibrium Au-Au spacing within DFT$+U$.
\end{abstract}

\maketitle

\section{Introduction}

Electron transport through atomic-sized metallic contacts in the low-bias regime has a ballistic nature and is commonly studied within the Landauer-B\"uttiker theory.
The current is carried by electronic quantum channels, which can be partly transmitted and partly reflected.
The ballistic conductance of the tip-nanocontact-tip system is proportional to the total transmission at the Fermi level, $G=\hgo\;T(\ef)$, where $e$ is the electron charge, $h$ is Planck's constant, and $T$ is the sum of the transmissions for the majority and the minority spin components.\cite{datta1995,agrait2003}
For an ideal one-dimensional conductor, such as a monoatomic metallic chain, the ballistic conductance is proportional to the number of bands crossing the Fermi level.\cite{agrait2003}
When the system contains some source of scattering, an adsorbed impurity or structural disorder, for instance, the conductance is lower than the number of bands, so that the transmission has to be explicitly computed by solving an electron scattering problem.\cite{datta1995}
In realistic systems, a convenient method for transmission calculations with density functional theory (DFT) and plane wave basis sets has been put forward by \citet{choi1999} in the case of Kleinman-Bylander pseudopotentials\cite{kleinman1982} (PPs), and has been extended to ultrasoft pseudopotentials\cite{vanderbilt1990} (US-PPs) by Smogunov and coworkers.\cite{smogunov2004,smogunov2004b}
This methodology has already been applied to several materials, such as monatomic chains of magnetic $3d$-transition metals,\cite{smogunov2004,smogunov2004b} tip-suspended chains of Ni,\cite{smogunov2006} Pd,\cite{gava2010} Pt,\cite{smogunov2008b}, and Au,\cite{sclauzero2012b} as well as to some impurity systems such as CO on Pt monatomic chains,\cite{sclauzero2008b} CO on Au chains\cite{sclauzero2012a} and nanocontacts\cite{sclauzero2012b} or atomic Ni on Au chains.\cite{miura2008}
The more widely used nonequilibrium Green's function technique (see Refs.~\onlinecite{Xue2002} and \onlinecite{Brandbyge2002}, for instance) has been applied to some of these systems (e.g., tip-suspended chains of Au\cite{Brandbyge2002} and Pt,\cite{strange2008} atop-adsorbed CO impurity on monatomic chains of Au\cite{strange2008} and Pt\cite{NengPing2011}), usually giving results in good agreement.

Conventional DFT-based methods correctly describe the ballistic transport properties of many systems,\citep{agrait2003} but are known to give wrong conductance values in weakly coupled molecular junctions (see, e.g., Refs.~\onlinecite{diventra2000,Toher2005PRL,thygesen2010}, and references therein).
This shortcoming of the DFT approach has been recently attributed to the self-interaction (SI) error,\cite{Toher2005PRL} which affects electron self-energies obtained through standard local or semi-local density functionals.\cite{perdew1981}
However, even in the class of systems where the standard exchange-correlation functionals usually yield good conductance values (e.g., metal nanocontacts and nanowires), SI errors might give rise to a wrong positioning of the conductor electronic bands with respect to the Fermi level and hence alter the number of available conductance channels.\citep{wierzbowska2005}
In Au monatomic chains, for instance, the $5d$-electron binding energies are lowered because of the SI so that $d$-bands are pushed toward the Fermi level (\ef) and two band-edges touch \ef\ for Au-Au spacings close to the equilibrium value.\cite{sclauzero2012a}
The spin-degeneracy of these bands is lifted by the Stoner instability induced by the extra density of states at \ef, resulting in a slightly magnetic ground state and two spurious conductance channels in addition to the two $s$-channels (taking into account spin).\cite{delin2003,miura2008} 
This theoretical prediction of the conductance thus gives $4\,\hgo$ for the pristine Au chain (or $6\,\hgo$ in the spin-unpolarized case\cite{sclauzero2012a}), while one would expect a value around $G_0=2\,\hgo$ from a single spin-degenerate channel, as shown by experiments on clean Au nanocontacts.\cite{agrait2003}

A rather simple, but very efficient and popular way to tackle the SI problem is the DFT$+U$ method, an extension to standard DFT originally aimed at improving the description of electron-electron correlations of strongly-localized electronic states (e.g., $3d$ states in transition metals or $4f$ states in rare earths).\cite{anisimov1991b,solovyev1994,liechtenstein1995,dudarev1998}
In this method, the Kohn-Sham (KS) Hamiltonian is augmented with a Hubbard-$U$ potential, which can be derived from a mean-field treatment of a many-body Hartree-Fock Hamiltonian acting on the manifold of localized orbitals.
Since SI is absent in the Hartree-Fock method, the SI error ascribed to the approximate functional will be partly relieved for the localized electron manifold.\cite{cococcioni2005}
A previous study on monatomic chains of $3d$ and $4d$ transition metals has shown that the DFT$+U$ method can improve the local density approximation (LDA) description of the electronic structure with an accuracy comparable to that of a more sophisticated SI correction scheme.\citep{wierzbowska2005}
A great advantage of DFT$+U$ is that it does not add any substantial complication to standard DFT techniques and retains their computational efficiency in treating very large systems, at variance with other methods more specifically designed to cope with SI errors, such as SI correction schemes\cite{perdew1981} or hybrid functionals.\cite{PBE0,HSE}

Some drawbacks or deficiencies, which have been only partly solved or addressed, can also be identified in this method.
For instance, the choice of the parameters in the Hubbard Hamiltonian (just $U$ in the simplest version, but many more if one considers magnetic exchange,\cite{solovyev1994,liechtenstein1995} spin-orbit coupling,\cite{bultmark2009} or inter-site interactions\cite{CampoJr2010}) may critically affect the results.
These parameters can be derived from renormalized atomic values, from constrained-occupation calculations (within linear muffin-tin orbital or similar methods\cite{anisimov1991a,solovyev1994,liechtenstein1995}), or through linear-response,\cite{cococcioni2005} but in several cases one or more set of values in a reasonable range are investigated.
The formulation of the Hamiltonian itself is subject to discussion because of the double counting term, which subtracts those energy terms that are already accounted for by the underlying density functional.\cite{bultmark2009}
The calculation of the on-site occupations also introduces some arbitrariness: they can be obtained by integrating the charge inside an atomic sphere,\cite{anisimov1991a} or derived from the overlap of the KS solutions with a set of localized wave functions\cite{cococcioni2005} (e.g., atomic-like states centered on the Hubbard atoms).
Nevertheless, the method has proven useful in many cases where standard density functionals fails and it is still widely used and still subject to development.\cite{bultmark2009,CampoJr2010}

In this work, we introduce a simple and efficient method to calculate the Landauer-B\"uttiker ballistic conductance within DFT$+U$ in a scheme based on plane wave basis sets and ultrasoft pseudopotentials.\cite{smogunov2004b,choi1999}
The DFT$+U$ on-site occupations are computed using atomic-derived wave functions truncated at the core radius of the corresponding PP,\cite{bengone2000} so that the Hubbard potential can be rewritten in terms of the PP projectors only and it can be incorporated in the non-local part of the PP as a simple renormalization of its coefficients.
We shall illustrate the method for the widely-used rotationally-invariant formulation of the Hubbard Hamiltonian,\cite{dudarev1998,cococcioni2005} but a generalization to more complex Hamiltonians would be straightforward.

We apply this scheme to study the effect of CO adsorption on the ballistic transport properties of an Au monatomic chain at the equilibrium Au-Au spacing.
Indeed, this system was previously investigated for different Au strains, but the low-strain limit could not be addressed because of the spurious contribution of $d$-electrons to the conductance.\cite{sclauzero2012a}
We show here that the Hubbard potential relieves the SI error of $5d$-states in the Au chain, so that the corresponding bands shift toward higher binding energies and do not present any band-edge at \ef.
As a consequence, a more realistic conductance of $1\,G_0$ and a non-magnetic ground state are recovered in the Au chain, thus allowing us to assess the effects of CO adsorption also in the low strain limit.
We find that the conductance reduction is substantially larger for the atop adsorption than for the bridge adsorption because of the different position of the transmission dip due to the hybridization resonance between the $5\sigma$-molecular level and the $s$-band of Au, much closer to \ef\ in the atop geometry.
This confirms the general picture given by the LDA study at larger strains.\cite{sclauzero2012a}

The paper is organized as follows: In \pref{sec:meth}, we provide the theoretical basis for embedding the Hubbard potential in the electronic structure and transport calculations through the pseudopotential coefficients.
In \pref{sec:goldchain}, we apply the method to the electronic, magnetic, and structural properties of the pristine Au chain.
Then, in \pref{sec:COatchain} we report the LDA$+U$ adsorption energetics of CO on the Au chain and we investigate the effect of the Hubbard $U$ on the transport properties. 
Finally, our conclusions will follow in \pref{sec:concl}.

\section{Method}\label{sec:meth}

\subsection{A scheme for ballistic transport within DFT$+U$}\label{sec:transplusU}

Since its earlier versions,\citep{anisimov1991a,anisimov1991b,liechtenstein1995} several variants of the DFT$+U$ method have been proposed.\cite{dudarev1998,bultmark2009,CampoJr2010} 
In order to illustrate our method, here we choose the simplified rotationally-invariant formulation of \citeauthor{dudarev1998},\cite{dudarev1998} which is widely used and has already been implemented in a plane wave-pseudopotential code.\cite{cococcioni2005,QE-2009}
In the rotationally-invariant Hubbard Hamiltonian, the main effect of the on-site Coulomb repulsion is described through a single parameter $U$ and the Hubbard energy term that is added to the usual DFT total energy reads as
\begin{equation}
  E_U = \frac{U}{2} \sis \smm \nmmp \left( \delta_{mm'} - \nmpm \right),
  \label{eq:EHub}
\end{equation}
where \nmmp\ are the on-site occupation matrices for atomic site $I$ and states of spin $\sigma$. 
The indexes $m$ and $m'$ run from $-l$ to $l$, where the orbital angular momentum $l$ is fixed by the choice of the localized electron manifold ($l=2$ for the $d$ shell, $l=3$ for the $f$ shell, \dots) and is omitted here for shortness of notation.
The occupation matrices can be written in a quite general form as:
\begin{equation}\label{eq:nmm}
	\nmmp = \sum_{\bfk v} f_{\bfk v} \sw{\pmmp}{\psi_{\bfk v}^{\sigma}} \;,
\end{equation}
where $\psi_{\bfk v}^{\sigma}$ are the solutions of the KS equation with spin $\sigma$, $f_{\bfk v}$ are  single-particle occupation factors, and \pmmp\ are generalized projection operators.\citep{cococcioni2005}
In plane wave codes, the latter can be conveniently chosen as fully separable,
\begin{equation}\label{eq:pmm}
  \pmmp = \ket{\fm}\bra{\fmp},
\end{equation}
where the \fm\ can be, for instance, atomic pseudo-wave functions. %
Consequently, the Hubbard potential
\begin{equation}
  \vu = \sum_I \smm \vmmp\, \pmmp\,, 
  \label{eq:VHub}
\end{equation}
will appear in the KS equation, with \vmmp\ defined as
\begin{equation}
  \vmmp = \frac{\de E_U}{\de \nmmp} = \frac{U}{2} ( \delta_{m'm} - 2\,\nmpm )\,.
  \label{eq:vmm}
\end{equation}
From \pref{eq:VHub} we notice that the Hubbard potential has the same form as the non-local part of a pseudopotential, which in the ultrasoft case is usually written as\cite{vanderbilt1990}
\begin{equation}
  V_{\textrm{NL}}^{\sigma} = \sum_I \sum_{ij} D_{ij}^{I,\sigma} \ket{\beta_i^I}\bra{\beta_j^I}\,,
  \label{eq:USPP}
\end{equation}
where $D_{ij}^{I,\sigma}$ are the screened coefficients and $\beta_i^I$ are the projector functions associated to atom $I$.
However, while PP projectors are nonzero only within a sphere of radius $r_{c}$ around the atom, the DFT$+U$ projectors in \pref{eq:pmm} have in principle an infinite support, even if in practice the atomic wavefunctions $\bk{\bfr}{\fm}$ decay rapidly to zero sufficiently far away from the atomic center $I$.
This feature of the Hubbard potential is not desirable because it makes the solution of the scattering problem more complicated and it would also increase its computational size.

This kind of problem can be overcome by limiting the integration for the on-site occupation matrices \nmmp\ within a sphere of finite radius centered on atom $I$, thus neglecting the contributions coming from the charge outside the spheres.
A similar approach has been proposed in the projector augmented-wave (PAW) framework by \citeauthor{bengone2000},\cite{bengone2000} who showed that the excluded charge is usually small for the localized states in the Hubbard Hamiltonian and that it is possible to use the PAW projectors instead of atomic wave functions in the Hubbard potential by adopting this approximation. 
The similarity between the US-PP method and the PAW formalism allows us to apply a PAW-like transformation\citep{bloechl1994} to the all-electron (AE) version of our projection operators,
$P_{mm'}^{\textsc{ae},I} = \ket{\phi_m^{\textsc{ae},I}}\bra{\phi_{m'}^{\textsc{ae},I}}$,
in order to obtain the corresponding pseudo (PS) version\footnote{%
This expansion holds only for ``quasilocal'' operators; for truly nonlocal operators there is an additional term, given by Eq. 12 of Ref. \onlinecite{bloechl1994}, which is neglected in this formula. Anyway it is formally zero for operators that vanish outside the spheres.} as in Ref.~\onlinecite{bengone2000}:
\begin{widetext}
\begin{equation}
  \begin{split}
    \sw{P_{mm'}^{\textsc{ae}}}{\psi_{\bfk v}^{\textsc{ae}}} = &\;\sw{P_{mm'}^{\textsc{ae}}}{\psi_{\bfk v}}\; + \sum_{ij} \bk{\psi_{\bfk v}}{\beta_i} 
    \left[ \mtx{\phi^{\textsc{ae}}_i}{P_{mm'}^{\textsc{ae}}}{\phi^{\textsc{ae}}_j} 
    - \mtx{\phi^{\textsc{ps}}_i}{P_{mm'}^{\textsc{ae}}}{\phi^{\textsc{ps}}_j}\right] \bk{\beta_j}{\psi_{\bfk v}}\raisetag{52pt}
  \end{split}\label{eq:pmmAE}
\end{equation}
\end{widetext}
where $\phi^{\textsc{ae}}_i$ ($\phi^{\textsc{ps}}_i$) is the AE (PS) partial wave corresponding to the PP projector $\beta_i$ of atom $I$, and $\psi_{\bfk v}$ are the PS wavefunctions that are obtained from the solution of the KS equation with US-PPs (we have omitted the atom index $I$ for shortness of notation).
The integrals needed for the AE and PS matrix elements between square brackets can be performed within the augmentation spheres around the selected atom ($|\bfr - \bfR_I| < r_c$), since AE and PS partial waves coincide outside the spheres by construction.

If we assume that the projection operators $P_{mm'}^{\textsc{ae}}$ are sufficiently localized within the atomic spheres and that partial waves and projectors form a complete basis inside those regions, we can apply to $P_{mm'}^{\textsc{ae}}$ the following equality:
\begin{equation}\label{eq:zeroPAWop}
  0 = \sw{\hat{B}}{\psi_{\bfk v}} - \sum_{ij} \bk{\psi_{\bfk v}}{\beta_i} \mtx{\phi^{\textsc{ps}}_i}{\hat{B}}{\phi^{\textsc{ps}}_j}\bk{\beta_j}{\psi_{\bfk v}},
\end{equation}
which strictly holds for any arbitrary operator $\hat{B}$ entirely localized within the atomic spheres.
This allows us to obtain from \pref{eq:pmmAE} an approximate expression for the projection operators:
\begin{equation}
	\sw{P_{mm'}^{\textsc{ae}}}{\psi_{\bfk v}^{\textsc{ae}}} \simeq \sum_{ij} \bk{\psi_{\bfk v}}{\beta_i} \mtx{\phi^{\textsc{ae}}_i}{P_{mm'}^{\textsc{ae}}}{\phi^{\textsc{ae}}_j} \bk{\beta_j}{\psi_{\bfk v}},
	\label{eq:pmmPS}
\end{equation}
where we neglected the contribution of the atomic wave functions outside the augmentation spheres.
If the above expression is used in \pref{eq:VHub}, the Hubbard potential can be rewritten in the following form:
\begin{equation}
	V_U^{\sigma} = \sum_I \sum_{ij} \left[\smm \vmmp \mtx{\phi^{\textsc{ae},I}_i}{P_{mm'}^{\textsc{ae},I}}{\phi^{\textsc{ae},I}_j} \right] \ket{\beta_i^I}\bra{\beta_j^I}\,,
	\label{eq:VHubPS}
\end{equation}
and it can be readily incorporated in the non-local part of the US-PP, resulting in 
\begin{equation}
	V_{\textrm{NL+U}}^{\sigma} \equiv V_{\textrm{NL}}^{\sigma} + V^{\sigma}_U = 
	\sum_I \sum_{ij} \left(D_{ij}^{I,\sigma} + \Delta_{ij}^{I,\sigma}\right) \ket{\beta_i^I}\bra{\beta_j^I}\;,
	\label{eq:vNLU}
\end{equation}
where $\Delta_{ij}^{I,\sigma}$ are the quantities between square brackets in \pref{eq:VHubPS}.
The effect of the Hubbard potential can thus be included in the KS equation through a simple renormalization of the non-local PP coefficients and no additional projectors other than those already required by the US-PP are needed.
The equation for the electron-scattering problem in the framework of US-PP and plane waves can be written as (Rydberg atomic units, $e^2/2=2m=h=1$, are used):\cite{smogunov2003,smogunov2006,choi1999}
\begin{equation}
  \big[ -\nabla^2 + V_{\textrm{eff}} + \hat{V}_{\textrm{NL}}' \big] \ket{\Psi_{\bfk}} = E \ket{\Psi_{\bfk}} \;,
  \label{eq:KSscat}
\end{equation}
where $\hat{V}_{\textrm{NL}}'$ is obtained from $\hat{V}_{\textrm{NL}}$ in \pref{eq:USPP} by replacing the screened coefficients with $\bar{D}_{ij}^{I,\sigma} = D_{ij}^{I,\sigma} - E\,q_{ij}^{I}$ (the $q_{ij}^{I}$'s being the integrals of the augmentation functions defined in Ref.~\onlinecite{vanderbilt1990}).
Therefore, the only additional step to include the Hubbard potential in this ballistic transport scheme is to replace $\hat{V}_{\textrm{NL}}'$ in \pref{eq:KSscat} with a new potential $\hat{V}_{\textrm{NL+U}}'$ where the coefficients are given by $D_{ij}^{I,\sigma}+\Delta_{ij}^{I,\sigma} - E\,q_{ij}^I$.
In this way, the DFT$+U$ transport calculation do not present any additional theoretical or technical difficulty than those already discussed and resolved in Refs.~\onlinecite{choi1999} and \onlinecite{smogunov2004b}.

\subsection{Computational details}

We have implemented the DFT$+U$ transport method outlined above in the \qe\ package,\cite{QE-2009} integrating it with the existing implementations of the DFT$+U$ method\cite{cococcioni2005} (in the \pwscf\ code) and of the ballistic transport with US-PP (\pwcond\ code\cite{smogunov2004b}).
The DFT$+U$ transmission calculations proceed in two steps: first, the ground-state electronic structure of the system is computed with \pwscf\ to obtain the local potential $V_{\textrm{eff}}$, the screened coefficients $D_{ij}^{I,\sigma}$, as well as the Hubbard coefficients $\Delta_{ij}^{I,\sigma}$.
In this step, the $\Delta_{ij}^{I,\sigma}$ must be updated at each iteration of the self-consistent loop, similarly to the US-PP screened coefficients, because of their dependency on the on-site occupations.
Second, the potential $\hat{V}_{\textrm{NL+U}}'$ is used to solve the scattering problem [\pref{eq:KSscat}] with the same techniques of Ref.~\onlinecite{smogunov2004b}.

In \pref{sec:goldchain}, we will apply this method to the isolated Au monatomic chain, while in \pref{sec:COatchain} we will study the ballistic transport through the chain in presence of a CO impurity adsorbed on it.
The calculations have been performed within the LDA using the same computational parameters described in Ref.~\onlinecite{sclauzero2012a} for the CO/Au chain system.
In particular, we use here the same US-PP of Au, C, and O, as well as the same plane wave cutoffs for the wave functions and the charge density presented there.
For the spin-polarized calculations of the infinite Au chain we reduced the smearing parameter for the electronic occupations to $0.005\,\ry$ and sampled the irreducible Brillouin zone with 46 $k$-points.
In our LDA$+U$ calculations, the Hubbard potential is applied to the $5d$-electron manifold of Au, either using the full atomic pseudo-wave functions as in \pref{eq:VHub}, or using the PP projectors as in \pref{eq:vNLU}.
For convenience, we named here ``atomic'' method the former and ``pseudo'' method the latter.
The electron transmission is calculated only for the ``pseudo'' method and the ballistic conductance has been obtained within the Landauer-B\"uttiker formalism by evaluating the total transmission at the Fermi energy, $G=e^2/h\cdot T(\ef)$.
We do not adopt here any self-consistent determination of $U$,\cite{cococcioni2005} but we rather compute the electronic structure for some values of $U$ in a range of interest.

\section{Electronic and magnetic properties of the Au chain}\label{sec:goldchain}

In this section we discuss the effects of the Hubbard potential on the structural, magnetic, and electronic properties of the infinite Au monatomic chain computed in the LDA.
We will compare the plain LDA results with the LDA$+U$ results obtained with both the atomic and pseudo methods (see above) for the on-site occupations.
Finally, we will validate the complex band structure (CBS) calculation for the pseudo LDA$+U$ method by showing that the usual band structure obtained with periodic boundary conditions (PBC) along $z$ is reproduced by the CBS at real $k_z$-points.

\begin{figure}[h!]
  \includegraphics[width=8.3cm]{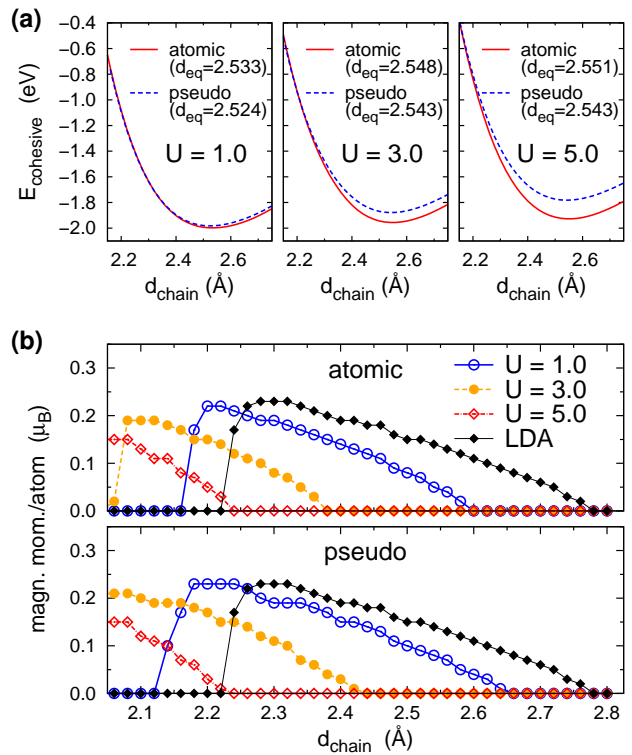}
  \caption{(Color online) Cohesive energy (a) and magnetic moment per atom (b) for the pristine Au chain as a function of the Au-Au spacing, \dwire.
  The LDA$+U$ results obtained with three different values of $U$ ($1.0\,\ev$, $3.0\,\ev$, and $5.0\,\ev$) are presented for each of the two projection methods (``atomic'' and ``pseudo'', see text).
  The corresponding equilibrium Au-Au spacing $d_{\mathrm{eq}}$ (in \ang) is reported in the insets of panel (a).
  In (b), the magnetic moments obtained from plain LDA calculations are also reported for comparison (black diamonds).}
    \label{fig:wireAuU}
\end{figure}

In \pref{fig:wireAuU}, we report the ground-state energy and magnetic moment per atom for the pristine infinite chain as a function of the Au-Au spacing, \dwire.
We study the dependence of these quantities on the strength of the Hubbard potential for several $U$ values up to $5\,\ev$.
For each value of $U$, the total energy of the isolated Au atom computed consistently is used as reference energy, hence the plots in \pref{fig:wireAuU}(a) represent the cohesive energy of the chain.
As $U$ is increased, the equilibrium spacing of the Au chain (see $d_{\textrm{eq}}$ in the insets) slightly expands from the plain LDA value ($2.51\,\ang$),\cite{sclauzero2012a} until it saturates to about $2.55\,\ang$ ($2.54\,\ang$) in the atomic (pseudo) method for $U \geqslant 3.0\,\ev$.
The cohesive energy instead decreases for both projection methods, but more rapidly for the pseudo method than for the atomic one.
The total ground state magnetic moment per atom for spacings $1.9\,\ang \leqslant \dwire \leqslant 2.8\,\ang$ is reported in \pref{fig:wireAuU}(b) for both pseudo and atomic methods.
The plain LDA result (diamonds) shows that for $2.22\,\ang < \dchain < 2.78\,\ang$ the ground state of the Au chain bears a finite magnetization.
The magnetic moment per atom reaches a  maximum value of $0.23\,\mu_{\textrm{B}}$ at $\dwire=2.3\,\ang$, while at the equilibrium spacing it is about 30\% smaller ($0.15\,\mu_{\textrm{B}}$).
When the Hubbard interaction is turned on, we observe a significant shift of the magnetic instability region toward smaller Au-Au spacings and a gradual reduction of the maximum magnetic moment as $U$ is increased.
Small quantitative differences between the two methods can be recognized, such as a more pronounced contraction of the magnetic instability region and of the maximum magnetic moment in the atomic method with respect to the pseudo method.
However, for strong enough Hubbard potentials ($U\geqslant3\,\ev$), both atomic and pseudo methods completely suppress the spurious magnetization of the Au chain at its equilibrium spacing.

\begin{figure}[tb]
  \includegraphics[angle=-90,width=8.3cm]{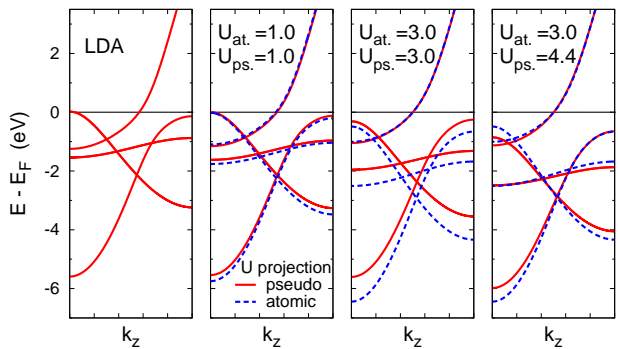}
  \caption{(Color online) Electronic band structure of the Au monatomic chain at the LDA equilibrium spacing for different values of the $U$ parameter (in \ev) and projection method (``atomic'' or ``pseudo'').
  In the leftmost panel the band structure from plain LDA is reported for comparison, while in the rightmost panel different values of $U$ have been used for the two projection methods.
  The bands are plotted for $k_z$ inside the irreducible Brillouin zone ($0 \leqslant k_z < \pi/a_3$, with $a_3=\dwire$).} 
    \label{fig:aubandsU}
\end{figure}

In \pref{fig:aubandsU}, we report the spin unpolarized electronic band structure of the Au chain obtained with plain LDA and with LDA$+U$ for selected values of $U$.
The large electron density of states at \ef, which is the main reason for the magnetic instability seen in the LDA, is due to the upper edge of the doubly-degenerate $\dxz/\dyz$ band pinned at \ef.
Within LDA$+U$, these bands, as well as the other filled bands (mainly of $d$-character), are progressively pushed toward higher binding energies as $U$ is increased, resulting in the suppression of the magnetic instability and of the spurious $d$-channels for $U$ values as small as $3\,\ev$. 
This effect is qualitatively the same for both atomic and pseudo methods, but the downward shift of the $d$-bands is more pronounced when using the full atomic wave functions.\footnote{This difference can be easily understood by noticing that the energy shift for a fully-occupied KS state having a perfect overlap with any of the Hubbard projectors is $-U/2$, which is the also upper bound for the shift.
This is almost the case in the monatomic chain when using full atomic projectors, while part of the charge is left out when using the PP projectors, resulting in smaller \nmpm\ values (about $10\%$ in our case) and hence in a weaker potential shift for the $d$-states.}
A perfect matching of the ``atomic'' and ``pseudo'' band structures is not possible because of the different broadening of the bandwidths, but a comparable energy shift can be obtained if a larger value of $U$ is used in the pseudo method (see rightmost panel of \pref{fig:aubandsU}).

\begin{figure}[bt]
  \includegraphics[width=8.3cm]{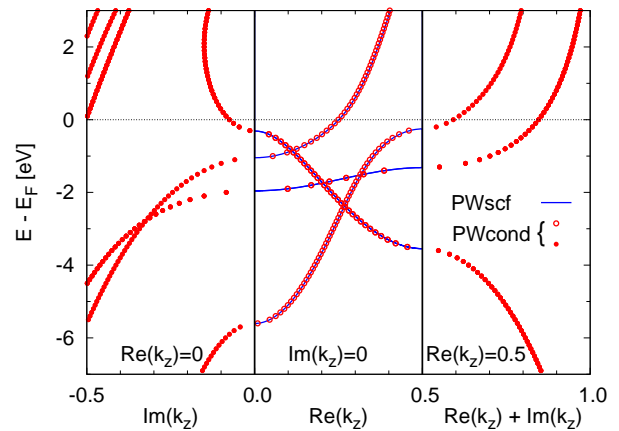}
  \caption{(Color online) Complex band structure of the Au monatomic chain within LDA$+U$ ($U=3\,\ev$) from \pwcond\ (circles and points) and band structure with full PBC from \pwscf\ (solid lines). 
  The central panel contains the CBS eigenvalues at real $k_z$ (circles), while those at imaginary $k_z$ and at $k_z=1/2+i\kappa$ are on the left and right panels (points), respectively (see also Ref.~\onlinecite{smogunov2004b}).
  The wave vector $k_z$ is expressed in units of $2\pi/a_3$, with $a_3=\dwire=2.51\,\ang$.}
    \label{fig:aucbsU}
\end{figure}

\begin{table*}[t!]
  \caption{\label{tab:geomhubbard}Optimized distances (in \ang) and chemisorption energies (in \ev) for the bridge and atop adsorption geometries within LDA and LDA$+U$.
The labels ``atomic'' and ``pseudo'' refer to on-site occupations computed using, respectively, atomic wavefunctions and PP projectors.
The reference energy of the isolated chain is consistently computed within LDA$+U$, using the same value of $U$ and type of projectors.}
  \begin{tabular}{l@{\hspace{2em}}ccc@{\hspace{2em}}ccc@{\hspace{3em}}ccc@{\hspace{2em}}ccc}
  \hline\hline
    & \multicolumn{6}{c}{Bridge}\hspace{3em} & \multicolumn{6}{c}{Atop} \\ 
    & \multicolumn{3}{c}{atomic}\hspace{2em} & \multicolumn{3}{c}{pseudo}\hspace{3em} &
    \multicolumn{3}{c}{atomic}\hspace{2em} & \multicolumn{3}{c}{pseudo} \\ 
 $U$&\dauc&\dco&\echem&\dauc&\dco&\echem&\dauc&\dco&\echem&\dauc&\dco&\echem\\ \hline
0.0\footnote{LDA results from Ref.~\onlinecite{sclauzero2012a}.} %
    & 1.952& 1.156 & $-$2.4&     &      &     & 1.891& 1.136 & $-$1.0&     &      &    \\ 
1.0& 1.957& 1.155& $-$2.4& 1.952& 1.156& $-$2.4& 1.901& 1.135& $-$1.0& 1.896& 1.135& $-$1.0\\
3.0& 1.971& 1.151& $-$2.1& 1.953& 1.154& $-$2.1& 1.931& 1.133& $-$0.8& 1.920& 1.134& $-$0.8\\
5.0& 1.988& 1.148& $-$1.7& 1.959& 1.151& $-$1.6& 1.969& 1.131& $-$0.6& 1.968& 1.133& $-$0.5\\
  \hline\hline
\end{tabular}
\end{table*}

In order to perform ballistic transport calculations within DFT$+U$, the CBS of the leads has to be computed at the same level of accuracy as the usual electronic band structure calculated using full PBC.
We test if our implementation meets this requirement by computing the CBS of the Au monatomic chain within LDA$+U$ for a representative value of $U=3\,\ev$.
In \pref{fig:aucbsU}, the CBS computed with the ballistic transport code \pwcond\ (symbols) is compared with the band structure obtained with the electronic structure code \pwscf\ (full lines). 
On the energy scale used in the figure, the CBS at real $k_z$ (central panel) is indistinguishable from the band structure with full PBC.
Therefore, when used in transport calculations this CBS will result in a single (spin-degenerate) transmission channel open at \ef\ for the Au chain, in more reasonable agreement with what is observed in experiments.

\section{Au chain with CO impurity}\label{sec:COatchain}

\subsection{Geometry and energetics}\label{sec:goldchaingeom}

We started by computing the optimized structures of the bridge and atop adsorption geometries of CO within LDA$+U$ for several values of $U$, following the same criteria described in Ref.~\onlinecite{sclauzero2012a}.
The molecule is kept in an upright position and the positions of C and O are optimized, while all Au atoms are kept fixed and aligned, equally spaced by the LDA equilibrium distance.

In \pref{tab:geomhubbard}, we report the optimized C-Au and C-O distances (\dauc\ and \dco, respectively) and the chemisorption energy (\echem) for the two geometries.
These results present a general trend of a progressive weakening of the interaction between the CO and the Au chain for increasing $U$ values in both geometries, as shown by the significant decrease of \echem, the increase of \dauc, and the shortening of \dco\ for $U>1\,\ev$.
Notice that the lowering of \echem\ due to the Hubbard $U$ correction does not change the energetic preference for the bridge site that was found in previous calculations.\cite{sclauzero2012a}
This conclusion does not depend on the projection method used, because the differences in \echem\ between the atomic and pseudo methods stay below $0.1\,\ev$, about one order of magnitude smaller than the bridge-atop \echem\ difference ($\sim1\,\ev$).

The weakening of the CO-Au interaction can be rationalized in terms of the energy shift of the Au $5d$ bands: larger values of $U$ will result into greater shifts of the $d$-band center toward higher binding energies, further and further away from the Fermi level.
This in turn will decrease the reactivity of the metal toward adsorbates, resulting in lower adsorption energies, larger adsorbate/metal binding distances, and smaller molecular bond relaxations.\cite{hammer1996}

\subsection{Ballistic conductance}\label{sec:goldchaintran}
\begin{figure*}[bt]
  \includegraphics[angle=-90,width=17cm]{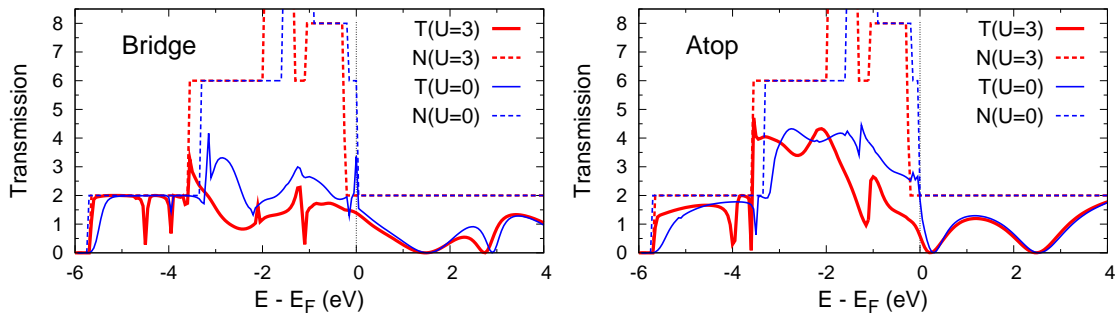}
  \caption{(Color online) Electron transmission (solid lines) and number of available transmission channels (dashed lines) as a function of the scattering energy for the bridge and atop geometries (left and right panels, respectively).
  The figure shows both the LDA$+U$ result with $U=3.0\,\ev$ (thick red lines) and the plain LDA result\cite{sclauzero2012a} (thin blue lines).}
    \label{fig:tranU}
\end{figure*}

In the previous section, we have shown that LDA$+U$ gives the correct number and type of conductance channels for the Au monatomic chain.
Here we will use this improved description of the CBS of the Au chain to study the effects of CO adsorption on the ballistic electron transport through the chain at low strains ($\dwire\simeq2.5\,\ang$ within LDA).
We shall first present the LDA$+U$ transmission for a selected value of $U$ and compare it with the plain LDA result,\cite{sclauzero2012a} then we will study the dependence of the tipless ballistic conductance on $U$.
The bridge and atop geometries obtained previously with plain LDA\cite{sclauzero2012a} will be used here, because we want to discuss the effect of $U$ through the electronic structure, rather than through the changes of the adsorption geometry (which are nevertheless small, see \pref{tab:geomhubbard}).

In \pref{fig:tranU}, the transmission through the Au chain with the CO impurity and the number of available channels (i.e., the transmission of the pristine infinite chain), are reported as a function of the electron scattering energy.
The figure displays both the LDA$+U$ results obtained with $U=3\,\ev$ and the previously obtained LDA results, reported here for comparison.
The removal of the spurious conductance channels due to the $\dxz/\dyz$ bands (cf.~\pref{fig:aubandsU}) gives rises to important changes in the ballistic conductance, which becomes $1\,G_0$ for the pristine chain within LDA$+U$, and drops below $1\,G_0$ in presence of CO.
The transmission of the remaining spin-degenerate $s$-channel is affected by CO in a rather different way in the two adsorption geometries and results in a much larger conductance for the bridge geometry ($G\simeq 0.7\,G_0$) compared to the atop geometry ($G\simeq 0.3\,G_0$).
The smaller conductance in the atop geometry stems from the dip in the $s$-transmission associated to the \sga-hybridization resonance,\cite{sclauzero2012a} which, at variance with the bridge geometry, falls very close to \ef.
This feature is also found in the LDA transmission, but the conductance-cutting effect of this interaction is hidden by the spurious conductance due to the extra $d$-channels at \ef.

At other scattering energies we also notice some changes in the transmission, especially below \ef, where the $d$-channels give the largest contribution.
In general, the LDA$+U$ transmission is slightly lower than the LDA one, excepting some energy regions (e.g. around $-3.5\,\ev$) where the number of channels increases in the LDA$+U$ because of the $d$-band shift to higher binding energies induced by the Hubbard potential.
Below $-3.5\,\ev$, a single $s$ channel is present and the main differences are localized in correspondence of transmission dips or at the lower band edge.
Above \ef, the number of transmission channels is not influenced by the Hubbard potential, nevertheless in the bridge geometry the transmission is modified between $2\,\ev$ and $3\,\ev$ because of a transmission dip shifting by about $0.2\,\ev$ toward \ef.
This dip is due to the antibonding \tps\ resonance,\cite{sclauzero2012a,sclauzero2008b} which is more coupled to the metal $d$-states with respect to the \sga\ resonance and is thus sensitive to the energy shifts of the $d$-bands. 

\begin{figure}[tb]
  \includegraphics[width=7.5cm]{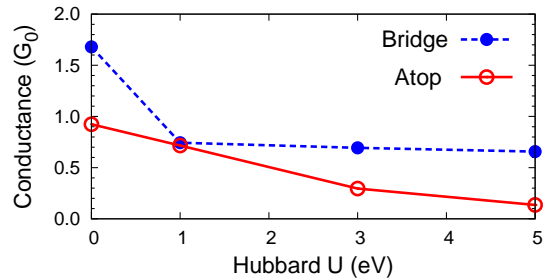}
  \caption{(Color online) Ballistic conductance in the bridge (blue points) and in the atop geometries (red circles) obtained for a selected set of $U$ values (lines connecting the calculated conductances have been drawn to guide the eye).} 
    \label{fig:condU}
\end{figure}

Finally, we discuss how the tipless ballistic conductance of the bridge and atop geometries depends on the strength of the Hubbard potential.
In \pref{fig:condU}, we report the conductance for some selected values of $U$ between $0\,\ev$ (plain LDA) and $5\,\ev$.
For small values of $U$, the tipless conductance decreases rather rapidly with $U$ because of the vanishing of the $d$-channel contributions to the conductance, and already at $U=1$ both geometries have a conductance lower than $1\,G_0$.
For $U\geqslant 1\,\ev$, the conductance as a function of $U$ stays almost constant in the bridge geometry, while it further decreases in the atop geometry as larger values of $U$ are considered.
However, for $U\geqslant 3\,\ev$ the two geometries have well-separated conductances, irrespectively of the value of $U$.
We can therefore conclude that the ballistic conductance in the bridge geometry is substantially higher than in the atop geometry not only at high strains, but also when the Au chain is not stretched.

\section{Conclusions}\label{sec:concl}

In conclusion, we have addressed the self-interaction problem in the electronic structure and in the ballistic transport through a simplified DFT$+U$ method, which leads to an efficient calculation of the electron transmission coefficients.
Using an approximate value for the on-site occupations entering the Hubbard Hamiltonian, we could include the effect of the Hubbard potential in the Kohn-Sham equation and in the scattering equation through a simple renormalization of the coefficients of the nonlocal pseudopotential.

We showed that the method is capable of suppressing the spurious magnetic instability of an Au monatomic chain arising at the equilibrium spacing in the LDA, thus recovering a non-magnetic ground state and removing at the same time the spurious contribution of Au $d$-states to the conductance.
We find that the chemisorption energies of CO on the Au chain can be substantially lower in the LDA$+U$ compared to the LDA, but the energetic preference for the bridge site previously obtained\cite{sclauzero2012a} with plain LDA is confirmed by the LDA$+U$ calculations.
The comparison of the ballistic conductance of the bridge and atop adsorption geometries, which is not feasible in the LDA without stretching the chain, reveals that the conductance cut mechanism seen at high Au strain takes place also at low strain.
Indeed, in the atop geometry the $s$-transmission dip due to the \sga\ resonance is very close to \ef\ and is responsible for the large conductance reduction, while in the bridge geometry the dip falls at higher energies and affects the conductance to a much smaller extent.

\begin{acknowledgments}
The authors are grateful to E. Tosatti and A. Smogunov for useful discussions.
G.S. also wishes to acknowledge G. Borghi for discussions about DFT$+U$.
This work has been supported by PRIN-COFIN 20087NX9Y7, as well as by INFM/CNR ``Iniziativa trasversale calcolo parallelo''. 
Computing resources have been provided by SISSA/Democritos eLab through its Linux cluster and by CINECA through the SISSA-CINECA agreement 2009-2010.
\end{acknowledgments}

%

\end{document}